\documentclass[reprint,
superscriptaddress,
amsmath,amssymb,aps,nofootinbib,
prb,
]{revtex4-2}

\usepackage{graphicx}
\usepackage{dcolumn}
\usepackage{bm}
\usepackage{color}
\usepackage{multirow}
\newcommand{\beginsupplement}{
  \setcounter{table}{0}  
  \renewcommand{\thetable}{S\arabic{table}} 
  \setcounter{figure}{0} 
  \renewcommand{\thefigure}{S\arabic{figure}}
}

\newcommand{\ECP}{EuCd$_2$P$_2$}

\begin{document}

\preprint{APS/123-QED}

\title{Robust Magnetic Polaron Percolation in the Antiferromagnetic CMR System EuCd$_2$P$_2$}

\author{Marvin Kopp}
 \email{kopp@physik.uni-frankfurt.de}
\author{Charu Garg}
\author{Sarah Krebber}
\author{Kristin Kliemt}
\author{Cornelius Krellner}
\affiliation{%
 Physikalisches Institut, Goethe-University Frankfurt\\
 Max-von-Laue Str.\ 1, 60438 Frankfurt am Main, Germany
}


\author{Sudhaman Balguri}
\author{Mira Mahendru}
\author{Fazel Tafti}
\affiliation{
 Department of Physics, Boston College\\
 140 Commonwealth Avenue, Chestnut Hill, MA 02467, USA
}%

\author{Theodore L. Breeze}
\affiliation{Department of Physics, Durham University, Durham,
DH1 3LE, United Kingdom}
\author{Nathan P. Bentley}
\affiliation{Department of Physics, Durham University, Durham,
DH1 3LE, United Kingdom}
\author{Francis L. Pratt}
\affiliation{ISIS Facility, STFC-Rutherford Appleton Laboratory, Harwell Science and Innovation Campus, Didcot, OX11 0QX, United Kingdom}
\author{Thomas J. Hicken}
\author{Hubertus Luetkens}
\author{Jonas A. Krieger}
\affiliation{PSI Center for Neutron and Muon Sciences CNM, 5232 Villigen PSI, Switzerland}
\author{Stephen J. Blundell}
\affiliation{Department of Physics, Oxford University, Clarendon Laboratory, Parks Road, Oxford, OX1 3PU, UK}
\author{Tom Lancaster}
\affiliation{Department of Physics, Durham University, Durham,
DH1 3LE, United Kingdom}

\author{M. Victoria Ale Crivillero}
\author{Steffen Wirth}
\affiliation{%
 Max-Planck-Institute for Chemical Physics of Solids\\
 N\"othnitzer Str.\ 40, 01187 Dresden, Germany
}%

\author{Jens Müller}
 \email{j.mueller@physik.uni-frankfurt.de}
\affiliation{%
 Physikalisches Institut, Goethe-University Frankfurt\\
 Max-von-Laue Str.\ 1, 60438 Frankfurt am Main, Germany
}%

\date{\today}

\begin{abstract}
Antiferromagnetic \ECP\ has attracted considerable attention due to its unconventional (magneto)transport properties. At a temperature $T_{\rm peak}$ significantly above the magnetic ordering temperature $T_\textrm{N} = 11$\,K a large peak in resistivity is observed which gets strongly suppressed in magnetic field, resulting in a colossal magnetoresistance (CMR), for which magnetic fluctuations and the formation of ferromagnetic clusters have been proposed as underlying mechanisms. Employing a selection of sensitive probes including fluctuation spectroscopy and third-harmonic resistance, Hall effect, AC susceptibility and $\mu$SR measurements, allows for a direct comparison of electronic and magnetic properties on multiple time scales. We find compelling evidence for the formation and percolation of magnetic polarons, which explains the CMR of the system. Large peaks in the weakly-nonlinear transport and the resistance noise power spectral density at zero magnetic field signify an inhomogeneous, percolating electronic system below $T^\ast \approx 2\,T_\textrm{N}$ with a percolation threshold at $T_{\rm peak}$. In magnetic fields, the onset of large negative MR in the paramagnetic regime occurs at a universal critical magnetization similar to ferromagnetic CMR materials. The size of the magnetic polarons at the percolation threshold is estimated to $\sim 1 - 2$\,nm. The mechanism of magntic cluster formation and percolation in \ECP\ appears to be rather robust despite large variations in carrier concentration and likely is relevant for other Eu-based antiferromagnetic CMR systems.    
\end{abstract}

\maketitle

\section{\label{sec:intro}Introduction}

The observation that the magnetic state of a system critically affects its electronic transport properties is at the heart of spintronics research and applications \cite{Chappert2007}.
One fundamentally important effect is the so-called colossal magnetoresistance (CMR), where the conductivity of materials drastically increases in a magnetic field, rendering such systems promising candidates for memory or sensor applications \cite{Marrows2016}. A wide variety of different material classes showing MR ratios ranging over of several orders of magnitude have been studied, among them europium chalcogenides, monoxide and hexaboride \cite{Molnar1967,Methfessel1968,Penney1972,Shapira1973,Suellow2000,Molnar_inbook_2007}, rare-earth perovskite manganites \cite{Helmolt1993,Jin1994,Millis1995,Solovyev1996,Millis1998,Coey1999,Dagotto2001,Salamon2001},
chromium spinels \cite{Lehmann1967,Lin2016}, or pyrochlores \cite{Ramirez1997,Majumdar1998}.
These are often complex materials with competing magnetic, elastic or orbital interactions leading to rich phase diagrams which may exhibit intrinsic (nonchemical) electronic phase separation resulting in transitions that are percolative in nature.
In fact, nanoscale electronic phase separation is suggested to play a critical role not only for the CMR effect \cite{Dagotto2001,Miao2020}, but also for high temperature superconductivity \cite{Kivelson2003} and in suppressing critical dynamics in quantum phase transitions \cite{Uemura2007}. Electronic phase separation therefore has been a subject of intensive recent theoretical and experimental interest, and magnetic field-dependent spatial inhomogeneities in the conductance of materials are thought to play a vital role in the magnetotransport effects in general \cite{Kagan2021}. 
The role of spatial inhomogeneities in the  magnetization, and a concomitant variation in conductivity, was 
first invoked in a discussion of defects in  antiferromagnetic (AFM) LaMnO$_3$. The defects were suggested to locally disturb global AFM order and produce local net ferromagnetic (FM) perturbations in the lattice \cite{deGennes1960}. 
If the defect carries charge, a magnetic polaron --- in analogy to the well-known dielectric polaron --- is created \cite{Kasuya1968}.
Experimental realizations, as well as detailed theoretical descriptions under varying conditions, have followed over the intervening years. 
Of relevance to the present discussion is the `large magnetic polaron', first experimentally discussed for the case of a diluted magnetic semiconductor \cite{Ohno1992} and theoretically described in Refs.\ \cite{Wolff1996,Kaminski2002}.
None of these early works, however, addressed the {\it dynamic} aspects of the problem central to this work: the percolative electronic phase transition that results from the overlap of spatial inhomogeneities in conductivity produced by magnetic polarons \cite{vonMolnar2001,Molnar_inbook_2007}. 

Experimentally, early indications for the concept of polaron formation were provided by magnetotransport measurements \cite{Molnar1967,Molnar1968}. In the context of the manifestation of the percolative transition in the magnetotransport a two-fluid model characterized by magnetization-dependent conductivities was formulated \cite{Salamon2001} in which, at high fields, the high conductivity branch percolates through the sample and dominates.
The percolative phase transition has been studied most extensively in the mixed-valence perovskite manganites and has been investigated also in EuB$_6$ as a far more simple model system. Methods like electron \cite{Uehara1999} and scanning tunneling microscopy \cite{Faeth1999,Renner2002,Pohlit_2018} gave an approximate extension of the polarons of order a few nanometers. For these systems, the theoretically predicted near divergence of $1/f$-type fluctuations seen in the resistance noise power spectral density upon approaching the percolation threshold has been experimentally verified in Refs.\ \cite{Podzorov2000,Das2012}.
Further important techniques comprise (small-angle) neutron scattering \cite{Teresa1997,Adams2000,Lynn2007} and muon-spin relaxation ($\mu$SR) \cite{Brooks2004}.

More recently, AFM materials have become of major interest for researchers as spintronic and quantum information technologies call for new materials without FM order \cite{Tokura2017, Smejkal2018, Baltz2018, Han2023}. In this context, various recently investigated europium compounds with an AFM ground state, e.g. Eu$_5$In$_2$Sb$_6$ \cite{Rosa2020,Crivillero2023}, Eu$_5$In$_2$As$_2$ \cite{Balguri2025}, Eu$M_2 X_2$ ($M=\textrm{Cd, Zn, Mn}$; $X=\textrm{As, P, Sb}$) \cite{Soh2018,Rahn2018,Ma2019,Jo2020,Wang2021,Wang2022,Sunko2023,Krebber2023,Homes2023,Li2024} or Eu$_{0.95}$Sm$_{0.05}$B$_6$ \cite{Crivillero2024},  have come to the forefront of research activities as some of them show a very large CMR effect.
Being considerably less studied than FM systems, AFM insulators exhibiting magnetic polaron formation, likely also of anisotropic nature, provide a rich playground for possibly new quantum states \cite{Crivillero2023}.\\
\begin{figure*}[t]
\includegraphics[width=\textwidth]{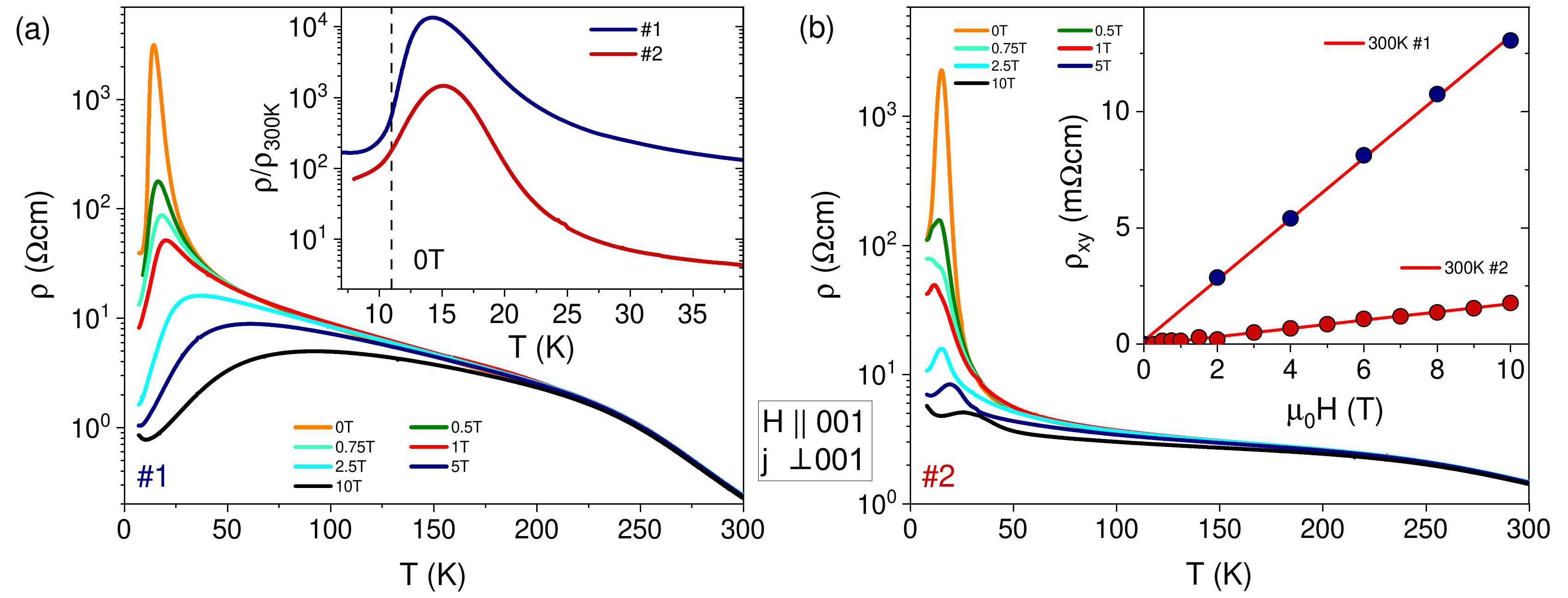}
\caption{\label{fig:resist} Comparison of the resistivities of sample \#1  and sample \#2  in (a) and (b), respectively, shown from $T=5\,$K to $T=300\,$K for magnetic fields from $\mu_0 H=0\,$T up to $\mu_0 H=10\,$T aligned along the $c$ axis.
Inset in (a) shows the normalized zero-field resistivities for both samples at low temperatures and the dotted line indicates $T_{\rm N}$.
Inset in (b) displays the measured Hall resistivities for both samples at $T=300\,$K. Lines are linear fits exhibiting a strong difference in slope for both samples in agreement with the differences in $\rho(300$\,K).}
\end{figure*}

In this work we focus on \ECP, a trigonal compound with an A-type AFM ground state (alternating FM layers with in-plane spins \cite{Wang2021}), where various reports in the past few years have discussed an unusual and pronounced peak in resistance, well above the N\'{e}el ordering temperature of $T_\textrm{N} \approx 11\,$K, which becomes strongly suppressed in an external magnetic field \cite{Wang2021,Sunko2023,Zhang2023,Usachov2024}.
Using a suite of sensitive probes of magnetism, Sunko \textit{et al.}\ discovered time-reversal symmetry breaking in \ECP\ at temperatures as high as $\sim 2\,T_\textrm{N}$ \cite{Sunko2023}. They proved its origin to be ferromagnetism, promoted by the magnetic exchange coupling between charge carriers, which become localized upon approaching the temperature of the resistivity peak, and the local Eu$^{2+}$ moments.
Their findings can be explained by the formation of FM clusters caused by a localized charge carrier aligning the spins of surrounding magnetic Eu$^{2+}$ ions in the extent of its wave function, i.e.\ by magnetic polarons.

The motivation of the present manuscript is twofold. First, we aim to test the hypothesis of FM cluster formation by providing more direct evidence for electronic phase separation and percolation of magnetic polarons using resistance noise spectroscopy and weakly-nonlinear transport measurements. We complement our findings with AC susceptibility, Hall effect and $\mu$SR measurements in order to provide a comprehensive picture of the correlations between magnetic and electronic properties.
Second, we aim to better understand how robust the picture of percolating magnetic polarons is in view of the strong sensitivity of the electronic properties to electronic impurities and charge carrier doping reported in the literature \cite{Wang2021,Sunko2023,Zhang2023,Chen_2024}. This is particularly important since it recently has been reported that the A-type AFM semiconductor with $T_\textrm{N} = 11$\,K exhibiting CMR can be transformed into a ferromagnet with $T_C = 47$\,K and metallic behavior by changing the growth conditions, apparently resulting in an increased charge-carrier concentration \cite{Chen_2024}.

\section{\label{sec:Results}Results}
\subsection{\label{subsec:Resistance}(Magneto)Resistance}
Resistivity measurements along the basal $a$-$a$ plane, with magnetic fields applied along the $c$-axis, were carried out using a standard four-terminal AC lock-in technique on two representative \ECP\ samples, one grown in the Frankfurt lab (sample \#1) and one in the Boston lab (\#2). The temperature-dependent resistivities for $T= 5 - 300\,$K are shown in Fig.\,\ref{fig:resist}(a) and (b) in magnetic fields up to $\mu_0 H=10\,$T for samples \#1 and \#2, respectively. In the inset of (a), the normalized zero-field resistivities $\rho(T)/\rho(300\,$K) are shown for temperatures below $T = 40\,$K. The overall qualitative behavior of $\rho(T)$ of the two samples is similar, exhibiting a semiconducting temperature dependence (${\rm d}\rho/{\rm d}T < 0$) upon cooling from 300\,K  down to a pronounced resistivity peak occurring at $T_{\rm peak} = T(\rho_{\rm max})$, a few K above the AFM ordering temperature $T_\textrm{N}$. For both samples we observe a distinct change of slope at about 240\,K.
However, there are a few remarkable quantitative differences. 
First, the room-temperature resistivity of sample \#1 of $\rho(300\,{\rm K}) = 0.23\,{\rm \Omega cm}$ \cite{Usachov2024} 
is significantly lower than the value of $1.4\,{\rm \Omega cm}$ of sample \#2. In comparison, the differences in the slopes of Hall resistivities measured at 300\,K shown in the inset of Fig.\,\ref{fig:resist}(b), which, in a simple one-band model, correspond to room-temperature carrier densities of $n_{\#1} = 4.8 \cdot 10^{17}\,$cm$^{-3}$ and $n_{\#2} = 3.5 \cdot 10^{18}\,$cm$^{-3}$. From this we can conclude that sample \#1 shows a much larger mobility at room temperature $\mu(300\,$K) (see Table\ \ref{table:ECP}). Importantly, both charge carrier concentrations are sufficiently low  to allow the formation of magnetic polarons, which we shall discuss in more detail below.\\ 
Second, the relative increase in resistivity from room temperature to $T_{\rm peak}$ amounts to $\rho(T_{\rm peak})/\rho(300\,{\rm K}) = 1.33 \cdot 10^{4}$ and $1.46 \cdot 10^{3}$ for \#1 and \#2, respectively, while the peak temperatures are $T_{\rm peak} = 14$\,K and 15\,K, in each case significantly above the AFM ordering temperature of $T_\textrm{N}= (11.0 \pm 0.2)\,$K for both samples, as determined from magnetization measurements. 
It is important to note that the semiconducting behavior observed for the present samples for $T > T_{\rm peak}$ is markedly different from the metallic behavior (${\rm d}\rho/{\rm d}T > 0$ for $T > 75$\,K) for the sample described in Refs.\ \cite{Wang2021,Sunko2023} with $T_{\rm peak} = 18$\,K and the same $T_\textrm{N} = 11$\,K.\\
Third, both samples show a saturation of the resistivity or even an upturn, in particular in high magnetic fields, at the lowest measured temperatures.

Fig.~\ref{fig:resist} also shows the resistivities of both samples in applied magnetic fields up to $\mu_0 H = 10$\,T along the $c$-axis, exhibiting the strongest suppression of more than four and three orders of magnitude for \#1 and \#2, respectively, for the highest applied field and at $T_{\rm peak}$. The quantitative differences between the samples become more obvious when plotting the usual magnetoresistance ${\rm MR} = [\rho(B) - \rho(0)]/\rho(0)$, see Fig.\ \ref{fig:kappa_MR}(a) in the Supplemental Information (SI).
For a relatively large field of $\mu_0 H = 5$\,T the maximum effect amounts to $-99.95\,\%$ and $-99.71\,\%$ for \#1 and \#2 at $T = 14$\,K and 16\,K, 
respectively. The MR sets in at higher temperatures for \#1, where at $\mu_0 H = 5$\,T and a temperature of 50\,K it already exceeds $-50\,\%$. 
For a relatively small field $\mu_0 H = 0.1$\,T, the onset of the negative MR at $T \sim 25 - 30$\,K is sharper and reaches values of about $-40\,\%$ for both samples.

\subsection{\label{sec:Noise} Resistance fluctuation spectroscopy and weakly-nonlinear transport}
\begin{figure}[]
\includegraphics[width=\columnwidth]{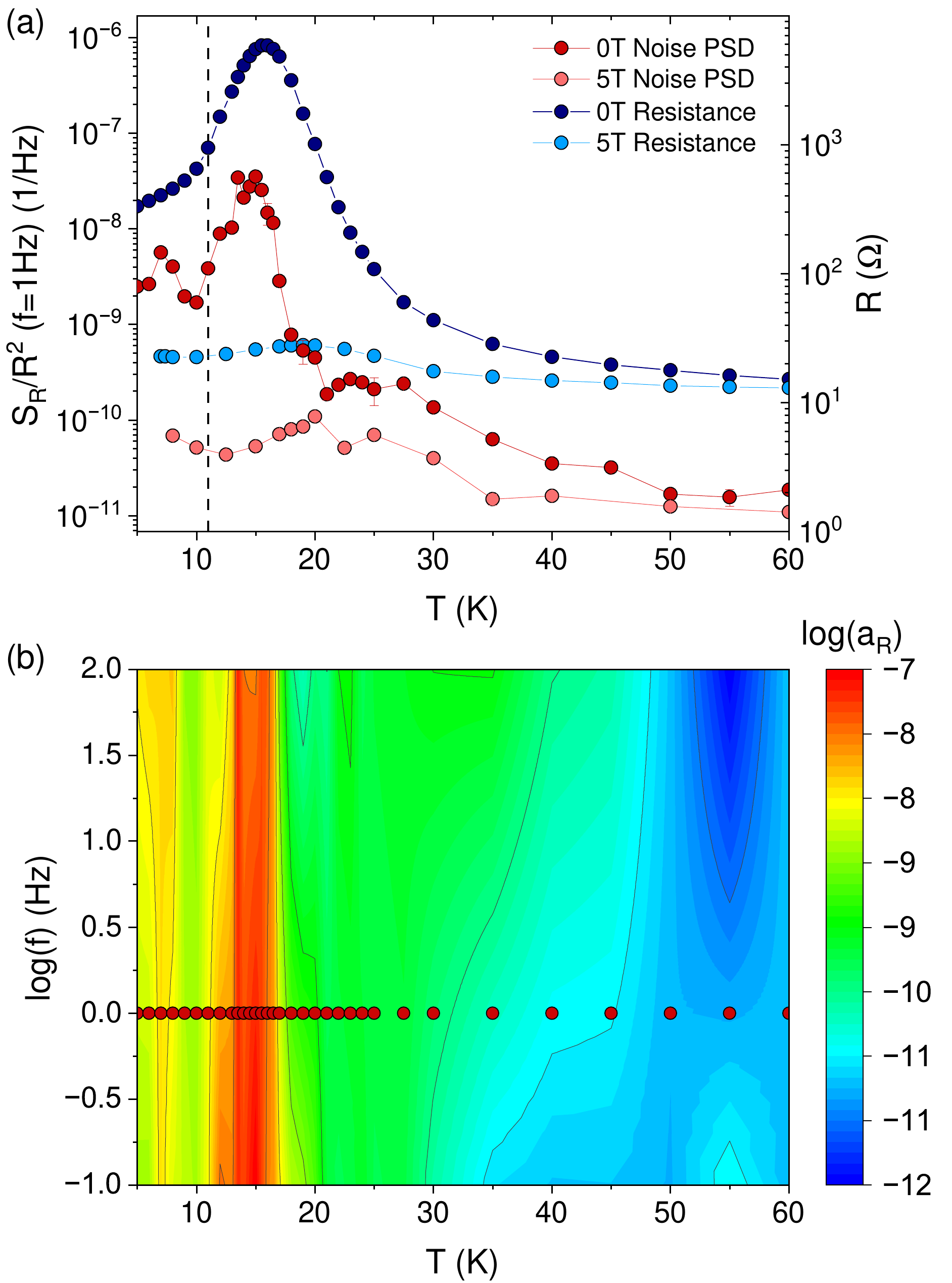}
\caption{\label{fig:noise}(a) Semilogarithmic plot of the 
normalized resistance noise power spectral density $S_R/R^2(T,f=1\,$Hz) (sample \#2) (reddish colors, left axis) and the corresponding resistance $R(T)$ (blueish colors, right axis). Dark red and dark blue denote the data in zero external field, the light colored data were measured in an external field of $\mu_0 H=5\,$T. (b) Noise map showing the relative amplitude $a_R = f \times S_R(f)/R^2$ in a logarithmic contour plot vs.\ temperature vs.\ $\log{{\rm frequency}}$ at zero magnetic field. The red dots correspond to the dark red data at $f = 1$\,Hz shown in (a). } 
\end{figure}
Measurements of the resistance noise power spectral density (PSD) $S_R(f,T)$ were carried out on sample \#2 for temperatures $T= 5 - 60\,$K. We observed generic $1/f^\alpha$ noise with a frequency exponent $\alpha = 0.7 - 1.4$. 
At certain temperatures Lorentzian spectra superimposed on the underlying $1/f$-type noise have been observed, a signature of two-level processes with 
finite life times.
By fitting the measured spectra at different temperatures with a superposition of $1/f$-type and Lorentzian noise PSD (see SI Fig.\ \ref{fig:Lorentz_eval}) 
we obtain the magnitude and corner frequency of the Lorentzian and the magnitude $S_R/R^2(T, f = 1\,\rm{Hz})$ and frequency exponent $\alpha(T)$ of the underlying $1/f$-type noise. Since the intermittently occurring two-level fluctuations showed no clear systematics in the measured temperature intervals, we focus in this work on the generic $1/f^\alpha$ noise. In Fig.\ \ref{fig:noise}(a) the noise magnitude $S_R/R^2(T,f = 1\,$Hz) at zero magnetic field (dark red), together with the corresponding resistance $R(T)$ (dark blue), are compared to their counterparts at $\mu_0 H = 5\,$T (noise: light red, resistance: light blue). 

Upon cooling down from $T=60\,$K the noise level at $f = 1$\,Hz slowly increases until about $T=30\,$K where it reaches a plateau even though the resistance continues to strongly increase. Note that the plateau region of constant noise level becomes wider in temperature with increasing frequencies, see Fig.\ \ref{fig:noise}(b), where the distribution of spectral weight is encoded in 
the color plot (on $\log$-scale) of the relative noise level $a_R = f \times S_R(f)/R^2 $ vs.\ temperature vs.\ $\log{\rm frequency}$, which is a dimensionless quantity characterizing the strength of the fluctuations.
At temperatures below the plateau the resistance noise PSD then steeply increases by more than two orders of magnitude until it reaches a maximum at $T \sim 13 - 15\,$K just below the resistance peak, before it drops to a minimum at about $T=10\,$K, the same temperature where the resistance changes slope (and shows an upturn in sample \#1). Upon further cooling, a smaller peak at about $T=7\,$K is observed. In comparison to the fluctuations at zero magnetic field, the measured low-frequency noise in $\mu_0 H = 5$\,T weakly increases upon cooling from 60\,K and becomes only weakly temperature dependent below 20\,K. While the negative MR increases gradually below about 60\,K, the suppression of the normalized resistance noise in magnetic field sets in rather abruptly below about 20\,K --- the temperature region where the CMR is largest [see Fig.\ \ref{fig:kappa_MR}(a)]. Compared to zero field, the noise PSD for  $\mu_0 H = 5$\,T is suppressed by $-99.98\,\%$ at 15\,K, which is essentially the same as for the MR. Thus, the mechanism responsible for the CMR is also responsible for the resistance fluctuations in this regime.\\ 
In basic percolation theory, often represented by the model of a simple random resistor network (RRN), weakly-nonlinear (third-harmonic) AC transport measurements probe microscopic inhomogeneities in the current distribution and are closely connected to the RRN's $1/f$-noise amplitude \cite{Yagil1992,Moshnyaga2009,Rammal1985} (see Methods part for details and SI for data of both samples). In Fig.\ \ref{fig:support}(a) the Fourier coefficient $\kappa_{3\omega}=V_{3\omega}/V_{1\omega}$ of the third-harmonic transport measured at $f=17\,$Hz is shown for $T=5 - 30\,$K for sample \#2 in different magnetic fields. In zero field $\kappa_{3\omega}$ starts to increase strongly below about $T^\ast \sim 22\,{\rm K} = 2\,T_{\rm N}$, coinciding with the increase of the resistance noise, and peaks at $T = 15.5\,$K. This increase and peak is strongly suppressed already in small magnetic fields until at $\mu_0 H=1\,$T $\kappa_{3\omega} \approx 0$ indicating a more uniform current distribution. The implications of this observation will be discussed below.

\subsection{\label{sec:mag} Magnetic properties}
\begin{figure}[t!]
\includegraphics[width=\columnwidth]{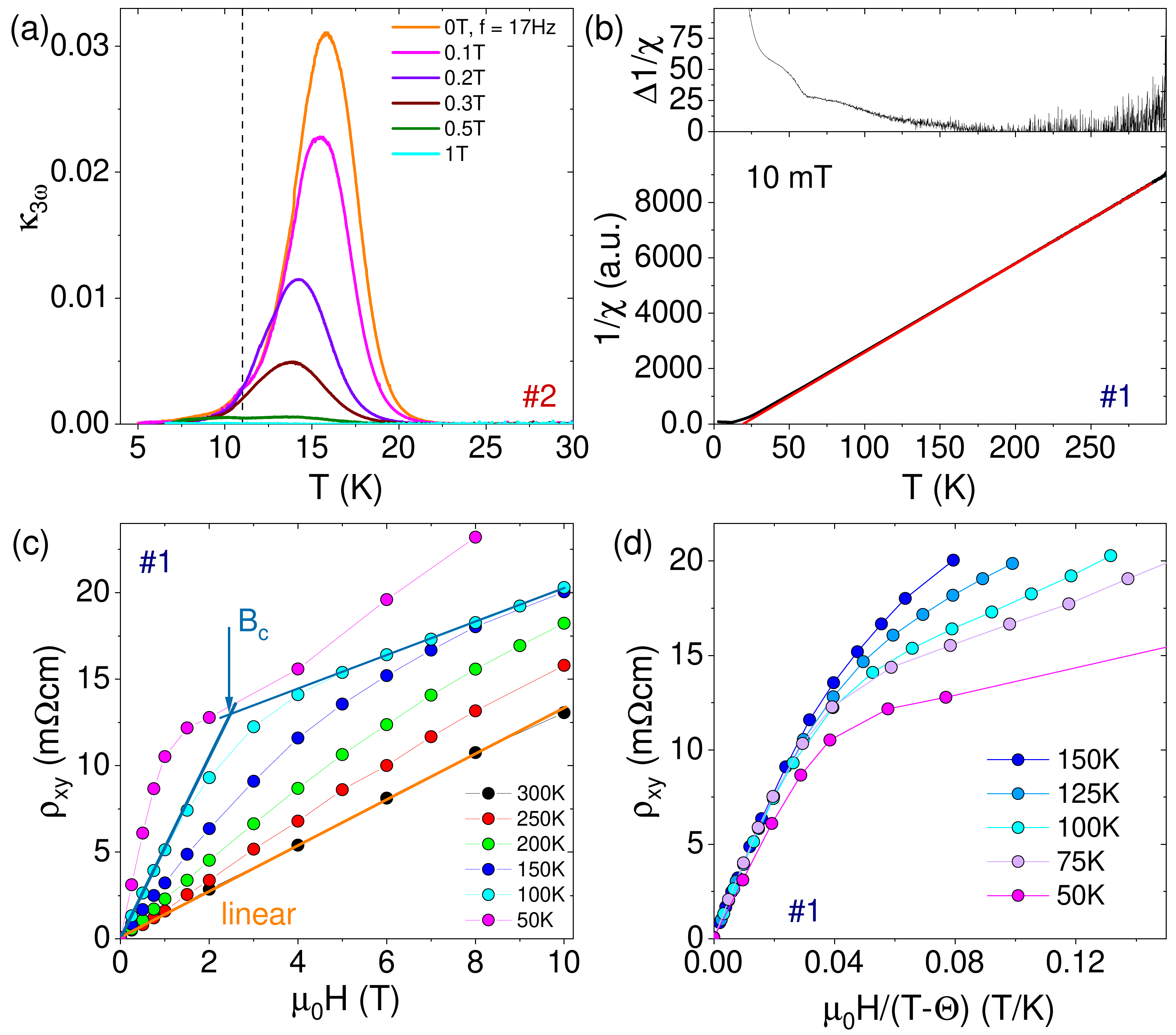}
\caption{\label{fig:support}(a) Fourier coefficient $\kappa_{3\omega}=V_{3\omega}/V_{1\omega}$ of the third-harmonic voltage vs.\ temperature in different magnetic fields for sample \#2. (b) Inverse DC magnetic susceptibility $1/\chi_{\rm dc}$ measured at $\mu_0 H=10\,$mT for sample \#1 with a linear fit to the data at high temperatures in red extrapolating to $\theta = 20$\,K. The upper graph shows deviations of the data from this linear fit on the same temperature scale. (c) Hall resistivity $\rho_{xy}$ measured at distinct temperatures for sample \# 1. The curve gradually deviates from a purely linear behavior at 300\,K (orange line) upon cooling. Blueish lines at $T=100\,$K represent linear slopes at small and large fields with a crossover field $B_c$. (d) Hall resistivity $\rho_{xy}$ for $T=150-50\,$K plotted versus the normalized field $\mu_0H/(T-\theta)$.} 
\end{figure}
The inverse magnetic DC susceptibility for sample \#1 for a field of 10\,mT displayed in 
Fig.\,\ref{fig:support}(b)
shows a Curie-Weiss behavior with an extrapolated paramagnetic Curie temperature of $\theta \approx 20\,$K (bottom panel) and deviates from a linear behavior below about 150\,K (top panel).

As shown in the inset of Fig.\,\ref{fig:resist}(b), the room temperature Hall resistivity $\rho_{xy}(B)$ shows a linear behavior from which we have extracted the carrier concentrations for both samples in a simple one-band model. However, curvature in $\rho_{xy}(B)$ develops gradually upon cooling, until two distinct slopes at $\mu_0 H = 0$ and at the maximum applied field of 10\,T can be fitted to the data determining a crossover field $B_c$ where the linear fits intersect, as exemplified for the data at 100\,K in Fig.\,\ref{fig:support}(c).  
Such behavior is reminiscent of the anomalous contribution to the Hall effect in ferromagnets. Upon further decreasing the temperature, the slope change becomes more pronounced and the crossover field becomes smaller, until below $T=50\,$K the Hall resistivity exhibits a shoulder-like feature related to $B_c$. A very similar behavior has been reported for EuCd$_2$As$_2$ where the anomalous contribution to the Hall effect was interpreted as the onset of quasi-static and quasi-long-range FM correlations \cite{Ma2019}. 
For  \ECP, in the same temperature regime where a clear curvature in the Hall resistivity develops the inverse magnetic susceptibility starts to deviate from linear behavior.\\
For the prototypical CMR system exhibiting magnetic polarons EuB$_6$, Zhang \textit{et al.} \cite{Zhang2009} have demonstrated a nonlinear Hall effect as a signature of electronic phase separation. Following their arguments, we show in Fig.\ \ref{fig:support}(d) the Hall resistivity $\rho_{xy}$ vs.\ a normalized field $\mu_0H/(T-\theta)$.
We find that for low fields the Hall resistivity curves collapse onto a single curve indicating that the transition at $B_c$ occurs at a single critical magnetization, very similar to the behavior of  prototypical EuB$_6$ and manganite CMR systems \cite{Zhang2009}. The observation that the onset of the MR for each magnetic field coincides with the temperature of the corresponding switching field $B_c$, see Fig.\ \ref{fig:alldata}(a) below,. This further underscores the significant role of magnetic polaron percolation for the CMR effect in \ECP.\\

\begin{figure}[b]
\includegraphics[width=70mm]{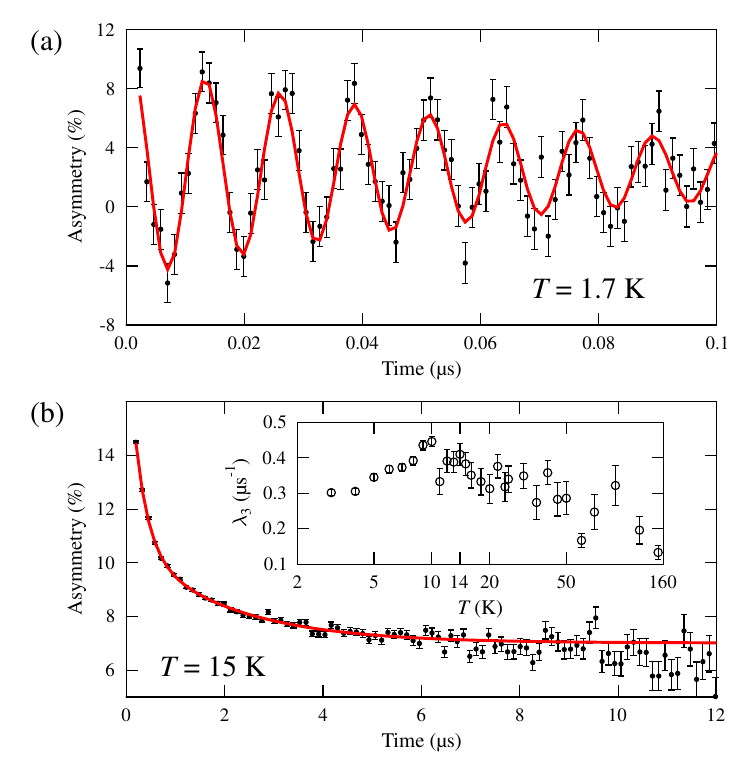}
\caption{(a) Example spectra from ZF $\mu$SR measurements made at (a) S$\mu$S and (b)  ISIS.
Inset: low relaxation rate $\lambda_{3}$ found in measurements made at ISIS. 
 \label{fig:ZF_ISIS}}
\end{figure}
In order to probe the local magnetic properties of the system and its dynamics we performed muon-spin relaxation ($\mu$SR) measurements, on samples grown similarly to the method described in Section~\ref{sec:methods} for sample \#1.
These measurements involve implanting spin-polarized muons in the material and measuring the subsequent muon-spin polarization, which is relaxed by the local magnetic field distribution at the muon sites. 
In zero-field (ZF) $\mu$SR measurements made at S$\mu$S  we observe high-frequency oscillations in the $\mu$SR spectra for $T<T_{\mathrm{N}}$, see Fig.\,\ref{fig:ZF_ISIS}(a), which result from muons experiencing a component of the local magnetic field transverse to the initial muon-spin direction. This behavior indicates that long-range magnetic order occurs in the sample at low $T$.

We fitted the spectra measured below $T=11$\,K with a single oscillatory component using the function
\begin{equation}\label{eqn:PSI_ZF_below}
A(t)=A_1{\rm e}^{-\lambda_1t}\cos 2\pi\nu t  + A_{\mathrm{bg}1},
\end{equation}
where $A_{\mathrm{bg}1}$ accounts for muons with their initial spin direction parallel to the local magnetic field, or those muons that stop in the sample holder. 
The frequency $\nu = \gamma_{\mu} \mu_0 H/2\pi$, where $\mu_0 H$ is the magnitude of the local field at the muon site and $\gamma_{\mu}$ is the muon gyromagnetic ratio, follows an order parameter-like decrease with increasing temperature, see Fig.~\ref{fig:alldata}(e).
A fit to the functional form $\nu(T) = \nu(0)[1-(T/T_{\mathrm{N}})^{\alpha}]^{\beta}$, see red line in Fig.\,\ref{fig:alldata}(e), yields $\alpha=1.3(1)$ and $\beta = 0.41(3)$, the latter being broadly consistent with fluctuations of Heisenberg moments in 3D close to the transition. We also estimate a  transition temperature $T_{\mathrm{N}}=11.2(1)$\,K, and ground state frequency $\nu(T=0)=85.6$\,MHz, corresponding to a local field of 0.63\,T.

Above $T_{\mathrm{N}}$, the spectra are found to relax exponentially, Fig.\,\ref{fig:ZF_ISIS}(b), typical for depolarization due to dynamics in the local magnetic-field distribution in the fast-fluctuation limit \cite{muonbook}. The zero-field spectra 
can be modeled using a function $A(t)=A_2{\rm e}^{-\lambda_{2} t} + A_{\mathrm{bg2}}$ with a relaxation rate $\lambda_{2}$ that rapidly decreases in the region $T_{\mathrm{N}}< T \leq T^{*}$ with a crossover temperature $T^{*}\approx 20 - 22$\,K. For $T > T^\ast$, the relaxation rate $\lambda_{2}$ remains at a roughly constant value around $2.5\,\mu$s$^{-1}$, see Fig.\,\ref{fig:alldata}(e). 

Complementary measurements were made at the ISIS muon source
which are better suited to following the evolution of the relaxation at high $T$. 

These reveal that the 2.5\,$\mu$s$^{-1}$ relaxation persists to  $T>160$\,K. An additional  slow relaxation, with  rate $\lambda_{3}$,  is observed at ISIS and shown inset in Fig.~\ref{fig:ZF_ISIS}(b), that we discuss below.
The limited time resolution at ISIS results in spectra that show a rapid change in relaxing amplitude $A_{2}$ in the temperature region $T_{\mathrm{N}}<T \lesssim 14$\,K (i.e.\ the $T$ range where the relaxation rate changes most rapidly in data measured at S$\mu$S). These data provide another probe of this temperature regime,  both in ZF and on the application of a longitudinal field of $\mu_0 H=100$\,mT.  The applied field is found to broaden this temperature regime, such that the temperature below which the amplitude starts to change increases to about $T^\ast$. This broadening is reminiscent of the analogous effect seen in the resistivity peak on application of a field, Fig.~\ref{fig:alldata}(e).

Finally, weak transverse field measurements were also made at ISIS that suggest that the non-magnetic signal does not vary appreciably across the temperature range. This implies that the phase separation that is suggested to occur in this system does not result in macroscopically-sized nonmagnetic regions whose volume varies with temperature, but rather is more likely to occur on the nanoscale such that it results in the temperature dependent relaxation captured by relaxation rate $\lambda_{2}$.

\section{Discussion}\label{Discussion}
\begin{figure}[t!]
\includegraphics[width=\columnwidth]{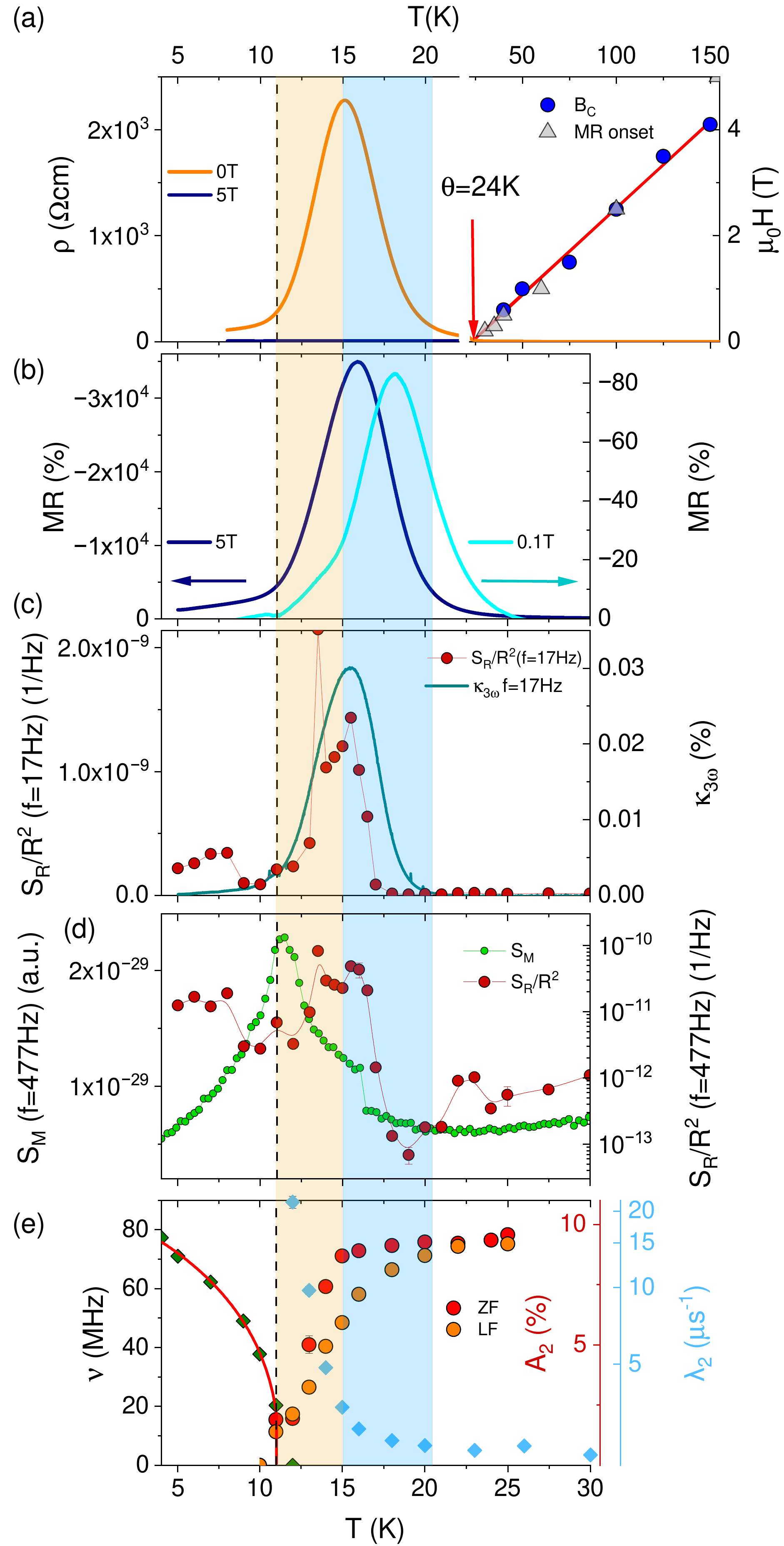}
\caption{\label{fig:data_combined}(a) Left: Resistance vs.\ temperature shown up to $2\, T_\textrm{N}$ in zero-field (orange) and $\mu_0 H=5\,$T (blue). Right: The crossover field $B_c$ of the Hall resistivity (blue) and onset of MR (grey triangles) are shown up to $T \sim 15\,T_\textrm{N}$, with a linear fit to the data. (b) Magnetoresistance in high and small field of $\mu_0 H=5\,$T and 0.1\,T (left and right axis), respectively. (c) Normalized resistance noise PSD, $S_R/R^2(f=17\,$Hz) on a linear scale in comparison to the third-harmonic Fourier coefficient $\kappa_{3\omega}$. (d)  Comparison of the calculated magnetic noise PSD $S_M(f=477\,$Hz) (green) and the resistance noise PSD $S_R/R^2(f=477\,\textrm{Hz})$ (red). (e) (i) Oscillation frequency $\nu$ (green diamonds), (ii) relaxation rate $\lambda_2$ (blue diamonds) and (iii) amplitude $A_2$ in zero-field (dark red dots) and in longitudinal field (orange dots), measured by $\mu$SR.} 
\label{fig:alldata}
\end{figure}

The main findings of the different methods are compiled in Fig.\,\ref{fig:data_combined}, where the magnetic transition is marked by the dotted line. Two important temperature regimes are marked by color: the orange area highlights the temperature range $T=11-15\,{\rm K}$ (between $T_\textrm{N}$ and $T_{\rm peak}$) and the blue one up to $T \sim 20$--$24\,{\rm K}\ (\sim T^\ast \approx 2\,T_\textrm{N}$). In (a) the resistance is shown on a linear scale in zero magnetic field (orange line) and $\mu_0 H=5\,$T (blue) for temperatures up to $2\,T_\textrm{N}$ (left). From $T = 2\,T_\textrm{N}$ up to $T = 150$\,K (right) the crossover field $B_c$ of the Hall resistivity is shown with a linear fit to the data, together with the onset of the MR (grey triangles). 
Despite the differences in carrier number of the present EuCd$_2$P$_2$ samples \#1 and \#2, their crossover fields $B_c$ behave strikingly similar, showing essentially linear behavior below about 150\,K  extrapolating to $T(B_c = 0) \approx 24\,$K, below which weakly-nonlinear transport and negative MR set in (see also Fig. \ref{fig:Hall} in SI). 

For reasons of comparison with the literature \cite{Wang2021,Zhang2023,Chen_2024}, in Fig.\,\ref{fig:data_combined}(b) we present the magnetoresistance alternatively calculated by
${\rm MR}=[\rho(B)-\rho(B=0\,\textrm{T})]/\rho(B)$
from the data in Fig.~\ref{fig:resist} for a large field of $\mu_0 H=5\,$T (dark blue) on the left axis and a small field $\mu_0 H=0.1\,$T (light blue) on the right axis. The onset of large MR roughly coincides with $T^\ast \sim 2\,T_{\rm N}$ and its maxima are in the blue-colored regime $T_{\rm peak} < T < T^\ast$.

The temperature dependence of the normalized resistance noise PSD, $S_R/R^2(T,f=17\,$Hz), is shown in Fig.\ \ref{fig:data_combined}(c) on a linear scale (red circles) together with the third-harmonic resistance coefficient $\kappa_{3\omega}$ (blue line) for zero magnetic field. Obviously, there is a large increase in both the resistance noise magnitude and the third-harmonic voltage generation, with the two effects roughly coinciding.
A strong noise and third-harmonic resistance peak was also observed in perovskite manganites \cite{Podzorov2000,Moshnyaga2009} and in EuB$_6$ \cite{Das2012}, and is interpreted as a hallmark of a microscopically inhomogeneous current distribution caused by the percolation of magnetic polarons at the temperature-/magnetic field-induced insulator-metal transition in these FM CMR systems. 
Although the present system is considerably more complex than one-component random resistor networks (RRNs), which often serve as simple model systems in percolation theory \cite{Rammal1985,Dubson_1989,Kogan1996}, this interpretation is based on the direct correlation between a strong increase in noise magnitude and a peak in weakly-nonlinear transport in a percolation scenario, i.e.\ when a conductive path through the sample is formed \cite{Rammal1985,Yagil1992,Moshnyaga2009}. This is corroborated by the characteristic power-law scaling behavior of the scaled weakly-nonlinear transport signal $V_{3 \omega}/I^3$ vs.\ $R$ with temperature as an implicit parameter, see Fig.~\ref{fig:perco} in the SI.    

Figure \ref{fig:data_combined}(d) shows the comparison between $1/f$-type magnetic noise PSD $S_M(T,f=477\,{\rm Hz})$ (linear scale, green circles) and resistance noise calculated for the same frequency (logarithmic scale, red) using the imaginary part of the AC susceptibility \cite{Kogan1996}:
\begin{equation}
\label{eq.:Sm}    
    S_M(f) = V \frac{2k_B T}{\pi f} \chi^{\prime \prime}(f),  
\end{equation}
where $V$ is the sample volume. 
Upon cooling below 60\,K, both types of fluctuations increase and exhibit a maximum at about 35\,K (see Fig.\,\ref{fig:magnet_AC} in the SI) which means that the magnetic noise in this temperature regime is caused by equilibrium fluctuations of the magnetization, and that the resistance and magnetization are strongly coupled. Upon further cooling, however, different noise sources determine the magnetic and resistance fluctuations. $S_M(f=477\,$Hz) peaks at the magnetic transition $T_{\rm N}$, whereas $S_R/R^2(f = 477\,{\rm Hz})$ exhibits a double peak structure in the temperature regime where the resistivity peak and the CMR are largest. 

In Fig.~\ref{fig:data_combined}(e) the main results of the $\mu$SR measurements, in part already discussed in section \ref{sec:mag} above, are displayed. From $\mu$SR, the system shows magnetic order below $T_{\mathrm{N}}=11$\,K. In the temperature regime $T_{\mathrm{N}}<T<T^{*}$ \ECP\ is not long-range magnetically ordered,  but there are dynamic fluctuations in the local magnetic field that 
rapidly relax the muon-spin polarization. 
Assuming a Redfield model for relaxation in the fast-fluctuation regime, we expect $\lambda_{2}\propto \gamma_{\mu}^{2}B_{\mathrm{a}}^{2}\tau$, where $B_{\mathrm{a}}$ is the amplitude of the fluctuating field. The relaxation rate of 20~$\mu$s$^{-1}$ seen just above $T_{\mathrm{N}}$ then corresponds to a fluctuation time of order $\tau\approx 0.1$\,ns. The rapid collapse of the relaxation rate for $T_{\mathrm{N}}<T<T^{*}$ implies that the cause of the rapid relaxation vanishes at $T>T^{*}$ and a distinct relaxation channel depolarizes the muons, with a characteristic fluctuation time an order of magnitude smaller.
It is notable that the characteristic temperature $T^{*}$ coincides with the onset of magnetic polaron percolation which implies that the same underlying electronic structure is responsible for both effects, despite their dependence on very different energy scales. 
In fact, a possible explanation of the muon response would be the occurrence of regions of slowly fluctuating magnetic moments locking together upon cooling below $T^{*}$, providing the broad spectral density of fluctuations required to relax the muons. 
We also note that, although an AFM model largely accounts for the muon response in the ordered regime, the measurements at ISIS feature evidence for an additional, slowly-relaxing background component, not resolved in the measurements made at S$\mu$S.
The fitted relaxation rate, $\lambda_{3}$, for this component is shown in the inset of Fig.\,\ref{fig:ZF_ISIS}(b), where we can see features that correlate with $T_{\mathrm{N}}$ and $T^{*}$. 
This contribution to the spectra likely results from muons stopping outside the sample in the silver  backing plate (which is absent at S$\mu$S) and might imply that the sample has a small FM response in its fluctuations, such that muons stopping outside the sample are relaxed parasitically.\\
\begin{table*}[t] 
    \centering
    \begin{tabular}{p{3cm}||p{2cm}|p{2cm}|p{2cm}|p{2cm}|p{2cm}} 
        Sample/Data origin& \#1 [this work] &\#2 [this work] & Wang \textit{et al.} \cite{Wang2021} & Zhang \textit{et al.} \cite{Zhang2023} & Chen \textit{et al.} \cite{Chen_2024} \\ 
        \hline
        $n_c$(300\,K) from Hall   & $4.8\cdot 10^{17}\,$cm$^{-3}$    & $3.5\cdot 10^{18}\,$cm$^{-3}$ &    $-6\cdot 10^{18}\,$cm$^{-3}$ & $\sim 1\cdot 10^{17} \newline \,$cm$^{-3}$ & $4.6\cdot 10^{19}\,$cm$^{-3}$ (at $150\,$K)\\
        \hline
        $\rho(300\,$K) & $2.3 \cdot 10^{-1}\, \Omega$cm & $1.4\, \Omega$cm & $2.4 \cdot 10^{-2}\, \Omega$cm & $3.1 \cdot 10^{-1}\, \Omega$cm & $5.5 \cdot 10^{-3}\, \Omega$cm \\
        \hline
        high-$T$ behavior & semiconducting & semiconducting & metallic & n.a. & metallic \\
        \hline
        $\rho(T_\textrm{peak})/\rho(300\,$K) & $1.33 \cdot 10^4$ & $1.46 \cdot 10^3$ & $5 \cdot 10^1$ & $5 \cdot 10^3$ & $0.87$ \\ 
        \hline
        $T_\textrm{peak}$ & $14\,$K & $15\,$K & $18\,$K & $14\,$K & $47\,$K \\
        \hline
        $T_\textrm{N}$ & $11\,$K & $11\,$K & $11\,$K & $10.9\,$K & $T_\textrm{C} = 47\,$K \\
        \hline
        ${\rm MR}_{\rm max}$ & $2.5\cdot 10^5\,$\% (14\,K, 5\,T) & $3.5\cdot 10^4\,$\% (16\,K, 5\,T) &  $5.6\cdot 10^3\,$\% (18\,K, 5\,T) & $1.75\cdot 10^5\,$\% (14\,K, 2\,T) & $80$\% (47\,K, 5\,T) \\
        \hline
        $\mu (300\, \textrm{K}) = (\rho e n_c)^{-1}$ & $56\,$cm$^2$/Vs & $1.3 \,$cm$^2$/Vs & $-43.3\,$cm$^2$/Vs &  $\sim200\,$cm$^2$/Vs & $24.6\,$cm$^2$/Vs \\
    \end{tabular}
    \caption{Properties of different \ECP\ samples. Zhang \textit{et al.} report sample-to-sample dependencies and some information is not available (n.a.).}
    \label{table:ECP}
\end{table*}

Based on these findings, we propose the formation and percolation of nanoscale FM clusters, {\it i.e.}\ magnetic polarons, to be responsible for the resistance peak at $T_\textrm{peak}$ and the strong CMR effect. 
Sunko \textit{et al.} \cite{Sunko2023} speculated that FM clusters begin to form at about 75\,K, where the resistivity of their sample shows a minimum, and undergo a merging/percolation transition at $T_{\rm peak}$. They were able to trace the signature of FM correlations up to $T \approx 2\,T_\textrm{N}$. Indeed, our findings of an enhanced $1/f$-type resistance noise PSD and the onset of a pronounced weakly-nonlinear transport signal (third-harmonic voltage) directly reveals dynamic fluctuations of the FM clusters and the presence of an inhomogeneous electronic system below $T^\ast$. The percolation threshold is identified as $T_{\rm peak}$, where the magnetic polarons overlap and form a conducting path through the sample which accounts for the strong decrease in resistivity. Furthermore, the extrapolation of the switching field $B_c$ in the Hall effect yields a similar temperature $T^\ast$, indicating stable magnetic polarons starting to percolate in zero external field. Finally, the amplitude and relaxation rate in $\mu$SR measurements deviates from their high-temperature behavior below $T^\ast$, indicating that the inhomogeneous system's fluctuations also occur on very fast timescales. Below $T_\textrm{N}$ the magnetic polarons freeze in the AFM matrix and continue to influence the magnetic and electronic properties of the system \cite{DawczakDebicki2024}.

While the onset of magnetic polaron percolation reveals clear signatures at $T^\ast \sim 2\,T_\textrm{N}$ in the noise and weakly-nonlinear transport, the temperature where magnetic polarons begin to form might be around $T=50\,$K, where the resistivity peak starts to develop and both the resistance and the magnetic $1/f$-type noise power spectral densities start to increase. This temperature range is comparable to Sunko \textit{et al.} \cite{Sunko2023}, where the crossover from high-temperature metallic to low-temperature semiconducting behavior (not observed for the present samples) is interpreted as the onset of FM cluster formation.

In comparison to FM EuB$_6$ \cite{Das2012}, the peak in the noise magnitude for \ECP\ is not only larger in amplitude but also broader. The magnetic noise $S_M$ obtained from measurements of the AC susceptibility is of $1/f$-type and shows a significant increase below $T \approx 100\,$K. Both $S_M$ and the resistance noise PSD reveal a broad peak (local maximum) between $T=50\,$K and $T^\ast$, see Fig.\ \ref{fig:magnet_AC} in the SI. Upon further cooling, both these slow fluctuations strongly increase whereas the peak in resistance noise coincides with the percolation threshold at $T_{\rm peak}$ and the peak in magnetic noise with $T_\textrm{N}$. The magnetic fluctuations appear to be driven by a competition between FM and AFM correlations, until the system orders antiferromagnetically. 

We now consider the model of correlated polarons by Bhatt {\it et al.}\ and Kaminski and Das Sarma \cite{Kaminski_2002,Bhatt2002}. For temperatures smaller than a characteristic energy scale given by the localization length of a dopant (captured hole or electron), all magnetic spins within distance $r_p(T) = \frac{1}{2} a_B \ln{(s S|J_0|/k_BT)}$ of dopants align forming a magnetically polarized cluster (polaron). Here, $a_B$ is the localization length of the donor, $s$ the carrier spin, $S$ the spin of the Eu$^{2+}$ in the matrix, and $J_0$ the effective exchange constant between the carrier and local moment spins. In Ref.\ \cite{Moshnyaga2009} it is discussed that the magnitude of the weakly-nonlinear transport signal $\kappa_{3 \omega}$ in a manganite thin film is directly related to the concentration of correlated polarons. With our finding of $\kappa_{3 \omega} \approx 3\,\%$ corresponding to the ratio of available carriers participating in magnetic polaron formation at $T_{\rm peak}$, and the observation of a $0.1\,$\% FM volume that stayed `frozen' at low temperatures in the ordered AFM phase of \ECP\ \cite{Sunko2023}, {\it i.e.}, the `infinite cluster', we can roughly estimate the size of magnetic polarons at $T_\textrm{peak}$. Assuming for simplicity randomly overlapping spherical bound magnetic polarons we find $r_p(T_{\rm peak}) \sim 1.3 - 2.2$\,nm for the two samples investigated here, a reasonable size of the critical percolation radius, which is also in line with the polaron size in EuB$_6$ \cite{Pohlit_2018}. Assuming a dielectric constant of $\epsilon = 30$ determining the localization radius $a_B$, then results in a magnetic exchange interaction between the charge carriers and the localized spins of $J_0 \sim 115 - 650\,$K. 
We note, however, that in the electronically and magnetically anisotropic compound \ECP\ the magnetic polarons are not expected to be spherical but rather shaped like an ellipsoid with the long axis along the in-plane easy magnetization direction \cite{Kagan2006,Ward2009,Gosh2022,DawczakDebicki2024}.

Finally, in Tab.\ \ref{table:ECP} we provide a comparison of various parameters relevant for the magnetotransport properties for different samples reported in the literature and in this work. Interestingly, samples exhibit both  metallic or semiconducting behavior upon cooling from room temperature and a significant range of charge carrier concentration $n_c$. Previous reports have established a correlation between higher charge densities and Eu vacancies, suggesting the possibility of intrinsic doping \cite{Chen_2024}. Samples reported in \cite{Chen_2024} are even metals with low $\rho(300\,{\rm K})$ and a positive coefficient ${\rm d}\rho/{\rm d}T > 0$ down to low temperatures where FM ordering occurs, highlighting the strong competition between FM and AFM interactions in this material. 
The magnitude of the CMR and also the ratio $\rho(T_{\rm peak})/\rho(300\,{\rm K})$ systematically increase with decreasing $n_c$, wheras $T_\textrm{peak}$ increases with increasing $n_c$.  
All samples that show a peak in the resistance due to a semiconducting behavior at low temperatures exhibit a CMR effect above the AFM ordering. Thus, magnetic polaron formation is a robust feature in \ECP\ and likely also in other members of this class of materials where the competition of AFM and FM interactions of localized spins and low charge carrier concentration gives rise to large magnetotransport effects.

\section{Methods}
\label{sec:methods}
\subsection{Sample growth}

Single crystals of \ECP\ have been succesfully grown from Sn flux, with two distinct samples resulting from slightly different growth processes provided by the institutions in Frankfurt (sample \#1) and Boston (sample \#2). The first sample (\#1) was grown using ingots of europium ($99.99\,$\%, Evochem), teardrops of cadmium ($99.9999\,$\%, Chempur), red phosphorous ($99.9999\,$\%, Chempur) and tin ($99.999\,$\%, Evochem) as starting materials. In the growth of sample \#2 sublimed ingots of europium (Alfa Aesar, $99.9\,$\%), cadmium tear drops (Alfa Aesar, $99.95\,$\%), red amorphous phosphorus powder (Alfa Aesar, $98.9\,$\%), and tin shots (Alfa Aesar, $99.999\,$\%) were used.\\
The elements were then cut into pieces and mixed together with a stoichiometry of $\textrm{Eu:Cd:P:Sn} = \textrm{1:2:2:20}$ under an inert Ar atmosphere inside a glove box. The materials were then placed in a crucible (\#1: graphite, \#2: alumina) inside an evacuated quartz ampule. In \#1 the elements were heated to $450\, ^\circ$C and held at this temperature for 5 hours. This ensured that the phosphorus gradually reacted with the other elements. Subsequently, the temperature was increased to $850\, ^\circ$C and held at this temperature for several hours in order to homogenise the melt. The temperature was then gradually decreased to $600\, ^\circ$C at a rate of 2\,K/h, where the Sn flux was removed by centrifugation, resulting in samples with a size of $3\,$mm$ \times 3\,$mm$ \times 1\,$mm. In the growth of the second sample (\#2), a similar procedure was followed, with the exception that the temperature was increased to $950\, ^\circ$C and held for 36 h. Following this, the temperature was then reduced to $550\, ^\circ$C at a rate of $3\,$K/h, where the flux was then removed by centrifugation, resulting in samples of similar size. 
The main difference between the two crystal growth methods are the choice of the inner crucible, the purity of the elements, and the temperature profile during the growth, which all could influence the overall quality of the crystals, in particular the concentration of intrinsic doping.

\subsection{Transport measurements}
Resistance measurements were carried out on two samples using a standard AC four point technique with a Lock in amplifier (e.g.\ SR830). To ensure good ohmic contacts, the sample surfaces were coated with thermally-evaporated gold ($200\,$nm) using a wetting layer of chromium ($7\,$nm) in a desired contact geometry, and then contacted with gold wires and conducting silver paste. The current was applied in the $a$-$a$ plane and magnetic fields always aligned to the $c$-axis. The third-harmonic resistance measurements were carried out in the same configuration. For the Hall measurements, the contact geometry with the smallest longitudinal component was used, and the voltages for discrete postive and negative field values have been antisymmetrisized.
Measurements of the noise power spectral density were carried out in a four-point configuration similar to a standard resistance measurement. Here the voltage measured from the sample is first amplified by a low noise amplifier (e.g.\ SR560) and the signal is then processed by a signal analyser (SR785) calculating the Fourier transformation and delivering the power spectral density
\begin{equation}
S_R(f) =  \lim\limits_{T \rightarrow \infty} \frac{1}{T}\Bigg|\int\limits_{-T/2}^{T/2} \delta R(t) \, \textrm{e}^{-2\pi ift} \, \textrm{d} t \Bigg|^2 \, .     
\end{equation}
A cross correlation with two voltage amplifiers and two lock-in amplifiers as described in \cite{Thyzel2024} was used to further reduce the noise background from the setup.

\subsection{Susceptibility}
DC susceptibility measurements were conducted using the vibrating sample magnetometry option of the Physical Property Measurement System (PPMS) by Quantum design.
Further magnetic measurements were conducted using a magnetic property measurement systems (MPMS3, Quantum Design Inc., San Diego, CA, USA). The magnetic AC susceptibility $\chi_\textrm{AC}$ was measured along the (001) axis ($c$-axis) with an applied AC field of 5 Oe after cooling in zero applied field (ZFC). We note that the magnetic properties of \ECP\ are highly sensitive to applied magnetic fields and therefore, the remnant field of the superconducting magnet was determined for each cool-down (typically 25 Oe) and compensated for. \\
The fluctuation-dissipation relations express the noise PSD in the equilibrium state of a system in terms of the dissipative part of the linear response of the same system. For comparison with the measured resistance noise which comprises a sum of different contributions we calculated the PSD of the magnetic fluctuations, $S_M(f)$, using the imaginary part of the AC susceptibility after Eq.\ \ref{eq.:Sm}.

\subsection{$\mu$SR}
A mosaic of single crystals of EuCd$_2$P$_2$ was mounted on a silver foil (foil thickness 25\,$\mu$m) with the $c$-axis out of the plane of the mosaic and parallel to the initial muon-spin direction. 
We made zero-field (ZF) $\mu$SR measurements \cite{muonbook} using the FLAME instrument at the S$\mu$S in the temperature range 1.5\,K to 30\,K using a $^{4}$He cryostat and ZF, longitudinal field (LF) and weak transverse field (wTF) measurements at ISIS using the EMU spectrometer. For the ISIS measurements, the sample was mounted on a Ag plate inside a $^{4}$He cryostat. 
Fitting functions for the S$\mu$S data are described in the main text. 
For ISIS ZF data, we used a fitting function
\begin{equation}\label{eqn:ISIS_ZF}
A(t)=A_2{\rm e}^{-\lambda_2t} +A_3{\rm e}^{-\lambda_3t} + A_{\mathrm{bg}3}.
\end{equation}
We found that $A_3$ could be held constant at two different values over these two regions.

We also completed supporting DFT calculations to find candidate muon stopping sites within the crystal using the MuFinder programme with the CASTEP code \cite{huddart2022mufinder,clark2005first}. A $2\times 2\times 2$ supercell was populated with a muon at a random position and the geometry was optimised to find a local minimum in energy, and this process was repeated 46 times. Calculations were done using the PBE functional with a $2\times 2\times 2$ k-point grid using a cutoff of 550\,eV. Muons were then moved to symmetry-equivalent positions and grouped based on position. A low energy site was found for muons aligned along the $c$ axis with the Cd and P ions, 0.5\,\AA\ above and below the $a$-$a$ plane of the Eu ions. We are able to calculate a local dipole field distribution experienced by muons at this position, from which muon spectra may be simulated. The AFM order at low temperatures contains FM layers in the $a$-$a$ plane which are antiferromagnetically coupled along the $c$ axis. From this ordering we constructed spectra which we found to be in good agreement with our experimental data for an Eu ordered moment of 5.4$\mu_{\mathrm{B}}$. Although this value is reduced compared to the full moment of the magnetic Eu ions ($7\mu_{\mathrm{B}}$), our calculation does not account for the hyperfine field at the muon site, which could account for some of the discrepancy.

\vspace{1cm}

\begin{acknowledgments}
We acknowledge support by the Deutsche Forschungsgemein– schaft (DFG, German Research Foundation) through Project 517733815 and SFB TRR288–422213477 (Projects B02 and A03).
The work at Boston College was funded by the U.S. Department of Energy, Office of Basic Energy Sciences, Division of Physical Behavior of Materials under award number DE-SC0023124.
Part of this work was carried out at the Swiss Muon Source (S$\mu$S), Paul Scherrer Institut, Switzerland,  and at the STFC-ISIS Facility, Rutherford Appleton Laboratory, UK and we are grateful for the allocation beamtime. NPB is grateful to Durham University for the provision of a studentship. This work is supported by EPSRC (UK). We also thank Dr.\ A.\ Virovets for the crystal analysis. Data taken in the Frankfurt and Dresden labs will be made available via DOI:XXXXXX. Data from the UK effort will be made available via DOI:XXXXXX.
\end{acknowledgments} 

\section*{Data availability}
All data are available in the main text or the supplementary materials.

\bibliography{EuCd2P2_paper}

\clearpage

\onecolumngrid

\beginsupplement

\section{Supplementary Information}

\subsection{Sample analysis}

\begin{figure}[h]
\includegraphics[width=0.5\columnwidth]{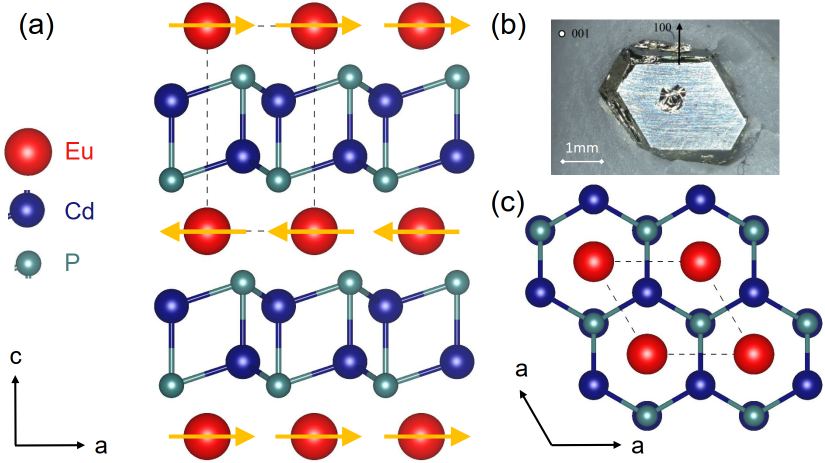}
\caption{\label{fig:structure}Structure of \ECP\ in (a) and (c) with A-type AFM order indicated by yellow arrows. (b) shows sample \#1 after polishing and before gold evaporation and contacting.}
\end{figure}

The structure and chemical composition of the \ECP\ samples was confirmed using powder X-ray diffraction (PXRD), energy dispersive X-ray spectroscopy (EDX) and single crystal analysis. The characterisation of the crystal structure of sample \#1 by PXRD yielded lattice parameters of $a = 4.324\,$\r{A}  and $c = 7.179\,$\r{A}. The powder diffraction patterns were recorded on a diffractometer with a Bragg-Brentano geometry and copper K$_\alpha$ radiation (Bruker D8).The chemical composition of sample \#1 was analysed with EDX, yielding averaged values of Eu$= (17\pm 2)$ at\%, Cd$ = (40 \pm 2)$ at\% and P $= (43 \pm 2)$ at\%. Analyis of sample \#2 yielded similar results within the error range.

\clearpage

\subsection{Weakly-nonlinear resistance and noise}

\begin{figure}[t!]
\includegraphics[width=1\textwidth]{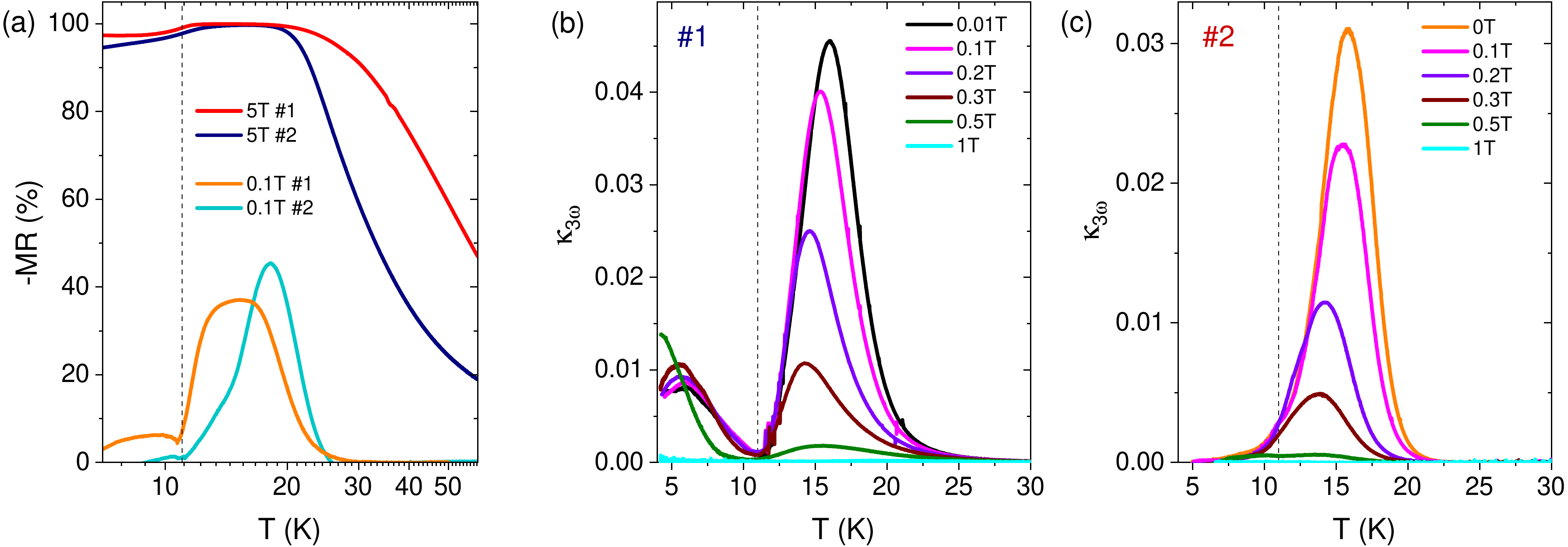}
\caption{\label{fig:kappa_MR}
(a) Negative magnetoresistance $-MR = -(\rho (B)-\rho (0))/\rho (0)$ at relatively small ($\mu_0 H = 0.1$\,T) and large fields $\mu_0 H=5\,$T for both samples \#1 (in orange and red) and \#2 (in light blue and blue). (b,c)
Fourier coefficient $\kappa_{3\omega}=V_{3\omega}/V_{1\omega}$ of the third-harmonic voltage generation below $T = 30\,$K for sample \#1 and \#2  measured in magnetic fields from $\mu_0 H=0\,$T to $1\,$T.}
\end{figure}

Besides the common resistance calculated by Ohm's law $R = V/I$, assuming a linear dependence between current $I$ flowing through, and voltage drop $V$ across the sample, in spatially inhomogeneous systems it is often useful to consider so-called higher harmonics of the voltage signal. Prime examples for intrinsic electronic and/or magnetic phase separation are the vortex lattice in a type-II superconductor, the magnetic skyrmion lattice, coexistence of phases near a first-order phase transition or the emergence of nano-scale clusters (e.g.\ polar nanoregions in relaxor ferroelectrics or magnetic polarons in CMR systems) \cite{Blatter_1994,Nagaosa2013,Thomas2024,Gennes1960,Kasuya1968,Das2012,Pohlit_2018}. The latter two examples are often discussed in the context of percolation in a random resistor network (RRN) \cite{Rammal1985,Stauffer2018}. Basic ideas of such a scenario are often invoked to describe the dynamics of magnetic polarons in CMR systems \cite{Gor’kov1999,Kaminski_2002}.\\ 
In a model percolation system of a semicontinuous metal film \cite{Yagil1992}, upon application of an AC current ($I = I_0 \cos(\omega t)$ the local resistance $r_\alpha$ can be written as 
\begin{equation}
r_\alpha = r_0 + \delta r_\alpha \cos(2\omega t + \phi)\, ,
\end{equation}
where $\delta r_\alpha$ are the fluctuations of each local resistance assumed to be significantly smaller than their resistance $r_0$ and $\phi$ is the phase shift between the heat production and the local temperature. 
The voltage across the sample can be written as
\begin{equation}
V = IR = I_0 R_0 \cos(\omega t) + \frac{1}{2} I_0 \Delta R \cos(3 \omega t + \phi) \, .
\end{equation}
Thus, a third-harmonic voltage 
\begin{equation}
V_{3f} \propto I_0 \Delta R \propto \frac{\sum i_\alpha^4 r_\alpha^2}{I_0}
\end{equation}
is created by the sample and connected to the fourth-order distribution of the local currents $i_\alpha$. 
In a simple RNN there is a direct connection to the PSD of the resistance fluctuations:
\begin{equation}
S_R/R^2 \propto \frac{V_{3f}}{I_0^3R^2}
\end{equation}
Here, energy conservation $I^2 \Delta R = \sum i_\alpha^2 \delta r_\alpha$ is used and a power-law scaling
\begin{equation}
\frac{V_{3f}}{I_0^3} \propto R^{2+w}
\end{equation}
is expected, where the critical exponent $w = \kappa/t$ depends on the particular percolation scenario \cite{Tremblay1986,Kogan1996}.
\\ 

The generated third-harmonic resistance or so-called weakly-nonlinear transport can be easily accessed in an AC lock-in measurement and reveals sensitive information on microscopic inhomogeneities in the current distribution of CMR systems \cite{Moshnyaga2009, Das2012}. The results are shown in Fig.\ \ref{fig:kappa_MR} for \#1 in (b) and for \#2 in (c). The Fourier coefficient $\kappa_{3 \omega} = V_{3 \omega}/V_{1 \omega}$ and is clearly nonzero above $T_{\rm N}$ and below about 25\,K and shows a pronounced peak for both samples at about $\sim 16$\,K in the vicinity of the resistance peak. We note that the observed values of $\kappa_{3 \omega} \sim 3 - 4\,\%$ in zero magnetic field are similarly large as the ones observed for the FM CMR systems EuB$_6$ (bulk) \cite{Das2012} and (La,Ca)MnO$_3$ (thin film) \cite{Moshnyaga2009}, where the third-harmonic generation is due to a percolative insulator-metal transition involving magnetic polarons.\\ 
Strikingly, the weakly-nonlinear transport peak is strongly suppressed already in small magnetic fields, with the peak shifting to lower temperatures with increasing field, and vanishes between about $\mu_0 H = 0.5$\,T and 1\,T for both samples. According to the common interpretation outlined above, this means a significant inhomogeneity in the microscopic current distribution in a rather well defined temperature interval above $T_{\rm N}$, which becomes suppressed in moderate magnetic fields. We note that the temperatures where $\kappa_{3 \omega} \gtrsim 0$ coincides with a finite negative MR at low fields and with the plateau-like regime of a large negative MR of more than 99\,\% at higher fields. Again, in agreement with the behavior of the MR at smaller fields, for sample 1, a second, smaller peak develops upon cooling below $T_{\rm N}$, the magnitude of which changes only slightly in increasing magnetic fields and is only suppressed in this temperature regime for fields of order 1\,T. Sample 2 does not show such a second peak and the weakly-nonlinear transport is close to zero below $T_{\rm N}$.\\

\begin{figure}[t!]
\includegraphics[width=1\columnwidth]{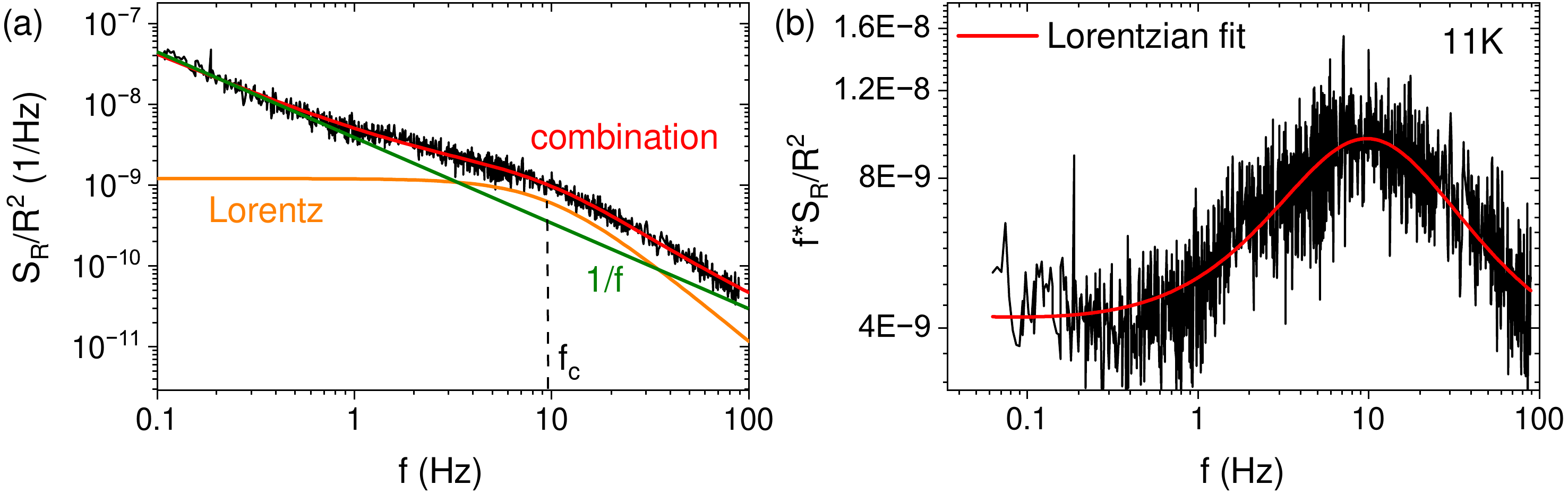}
\caption{\label{fig:Lorentz_eval}Typical PSD spectrum $S_R/R^2(f)$ of a $1/f-$type noise (green line) superimposed with a dominating two-level fluctuator producing a Lorentz spectrum (orange) in (a). Plot of $S_R/R^2 \cdot f$ with a fit (red) to evaluate the amplitude of the $1/f$-background, it's frequency exponent $\alpha$, the amplitude of the lorentzian and the corner frequency $f_c$ (b).}
\end{figure}

An examplary measured power spectral density (PSD) $S_R/R^2$ of the resistance fluctuations is shown in Fig.\ \ref{fig:Lorentz_eval} (a) in a double logarithmic plot, where the Lorentzian contribution is shown in blue, the linear $1/f$ background in red and the resulting combination in orange, describing the measured spectrum in black. Also marked is the corner frequency $f_c$ of the Lorentzian spectrum, which is connected to the lifetimes of the two-level fluctuator by $f_c = 1/(2 \pi) \cdot (1/\tau_1 = 1/\tau_2)$. By plotting $S_R/R^2 \cdot f$ as shown in (b), one can fit the spectrum with a Lorentzian function and obtain the amplitude and slope of the $1/f$ contribution, as well as the corner frequency $f_c$ and the amplitude of the lorentzian contribution. \\

\begin{figure}[h!]
\includegraphics[width=1\columnwidth]{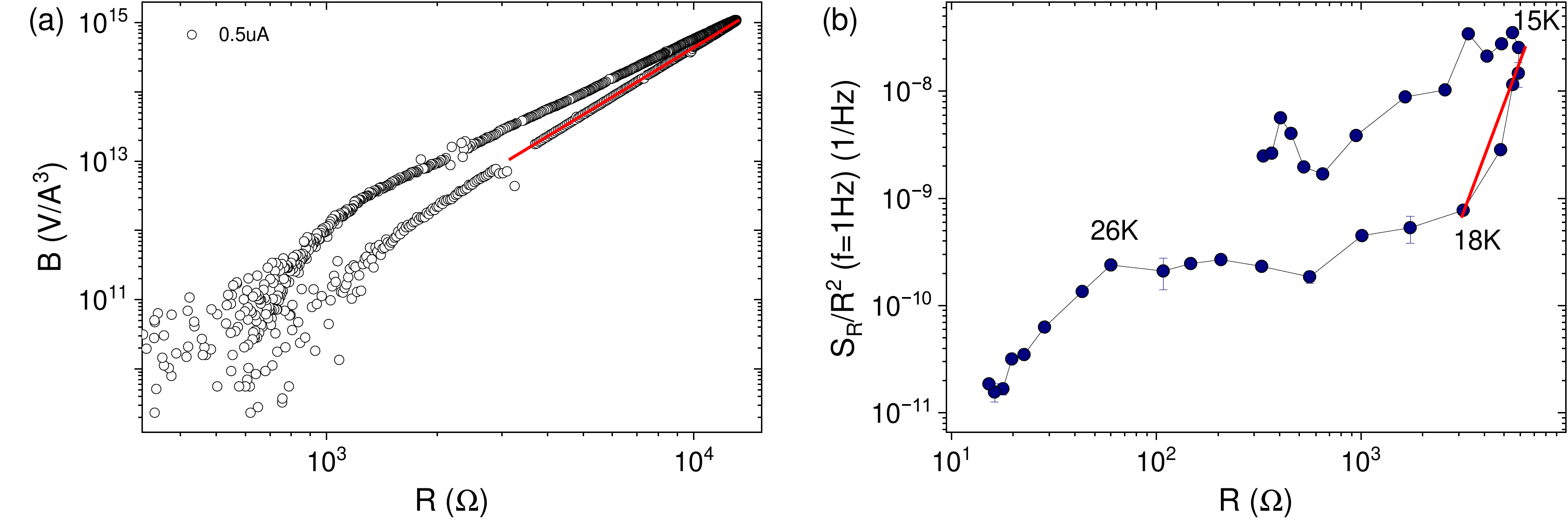}
\caption{\label{fig:perco}Plot of the normalized third-harmonic amplitude $B(T)=V_{3\omega}(T)/I^3$ versus the linear resistance $R(T)$ on a double logarithmic scale (a). The data can be linearly fitted between $T=18\,$K and $T=15\,$K yielding $\omega = 1.24$. A fit in the same temperature range for $S_R/R^2 (f=1\,$Hz) plotted versus $R(T)$ (b) yields a larger $\omega =4.4$ but also with a much larger error $\pm 0.7$.}
\end{figure}

In a percolation scenario a power-law dependence $B = V_{3\omega}/I^3 \propto R^{2+\omega}$ should be observed for the weakly nonlinear resistance \cite{Yagil1992}. As shown in Fig. \ref{fig:perco} the data can be linearly fitted between $T=18\,$K and $T=15\,$K yielding $\omega = 1.24$ very similar to the observation in silver thin films (or $\omega = 0.53$ for $T<15\,K$). A fit in the same temperature range for $S_R/R^2 (f=1\,$Hz) plotted versus $R(T)$ yields a larger $w=4.4$ but also with a much larger error $\pm 0.7$ as the fit is worse and there are less points to average over (or $\omega = 1.18 \pm 0.47$ below $T=15\,K$). In theory the scaling should also be $S_R/R^2 \propto R^{\omega}$.
It should be mentioned, that scaling of the third-harmonic signal with current and frequency, as well as basic scaling relations of the noise PSD expected for simple RRNs do not yield the same results for the present system, but vary with the chosen temperature range, which likely is due to the intricate interplay between competing AFM and FM interactions in \ECP\ (see Fig. \ref{fig:perco}).

\clearpage

\subsection{Magnetic and magnetotransport measurements}

\begin{figure*}[t]
\includegraphics[width=\textwidth]{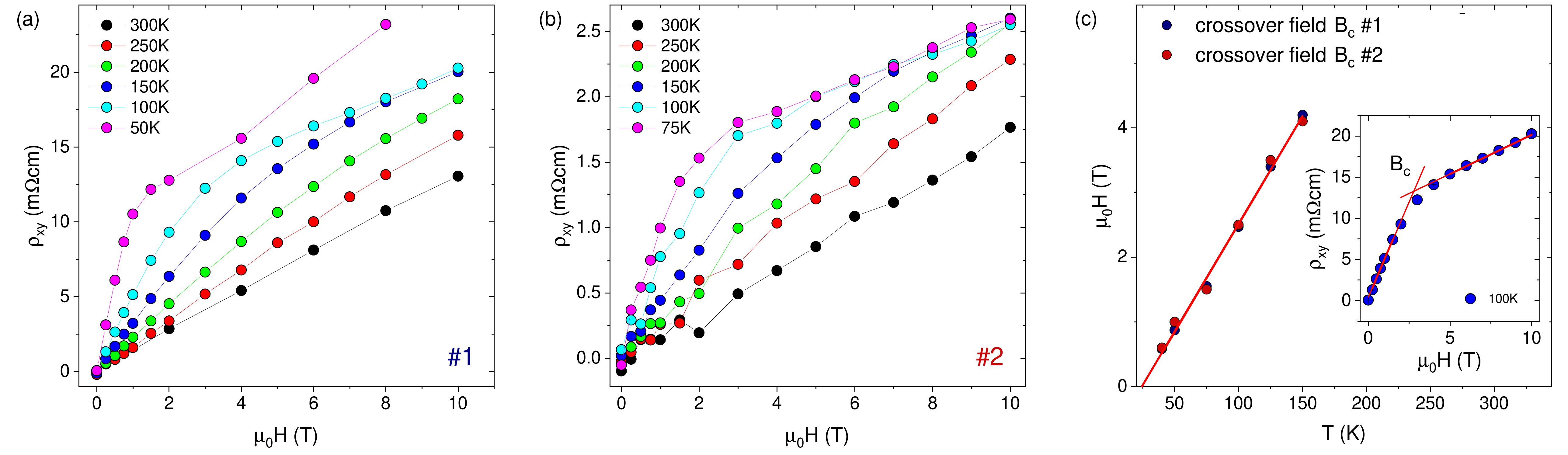}
\caption{\label{fig:Hall}(a) Hall resistivity $\rho_{xy}$ measured at distinct temperatures are shown for sample \#1 and \#2 in (b). The linear behavior at high temperatures develops into a curved shape upon cooling with two distinct slopes at low and high magnetic fields. Lines are guides to the eyes.
(c) Crossover field $B_c$ where the two slopes intersect, shown for both samples \#1 and \#2 in blue and red, respectively. Inset shows the construction at $T=100\,$K.}
\end{figure*}

Hall measurements were carried out on the same samples \#1 and \#2 using the given contact geometry. In order to cancel out the magnetoresistance offset due to the non-perfect Hall cross geometry, voltages have been measured at discrete positive and negative magnetic fields antisymmetrized.

Figure \ref{fig:Hall}(a) displays the Hall resistivity $\rho_{xy}$ vs.\ $\mu_0 H$ up to 10\,T for various discrete temperatures from $T = 300$\,K down to 50\,K for sample \%1, see SI Fig. \ref{fig:Hall} for the data of \#2. The temperature profiles for both samples are rather similar, despite the about one order of magnitude larger slope at room temperature (and correspondingly lower carrier concentration) of \#1, see inset of Fig.\ \ref{fig:resist}(b).

\begin{figure*}[t]
\includegraphics[width=\textwidth]{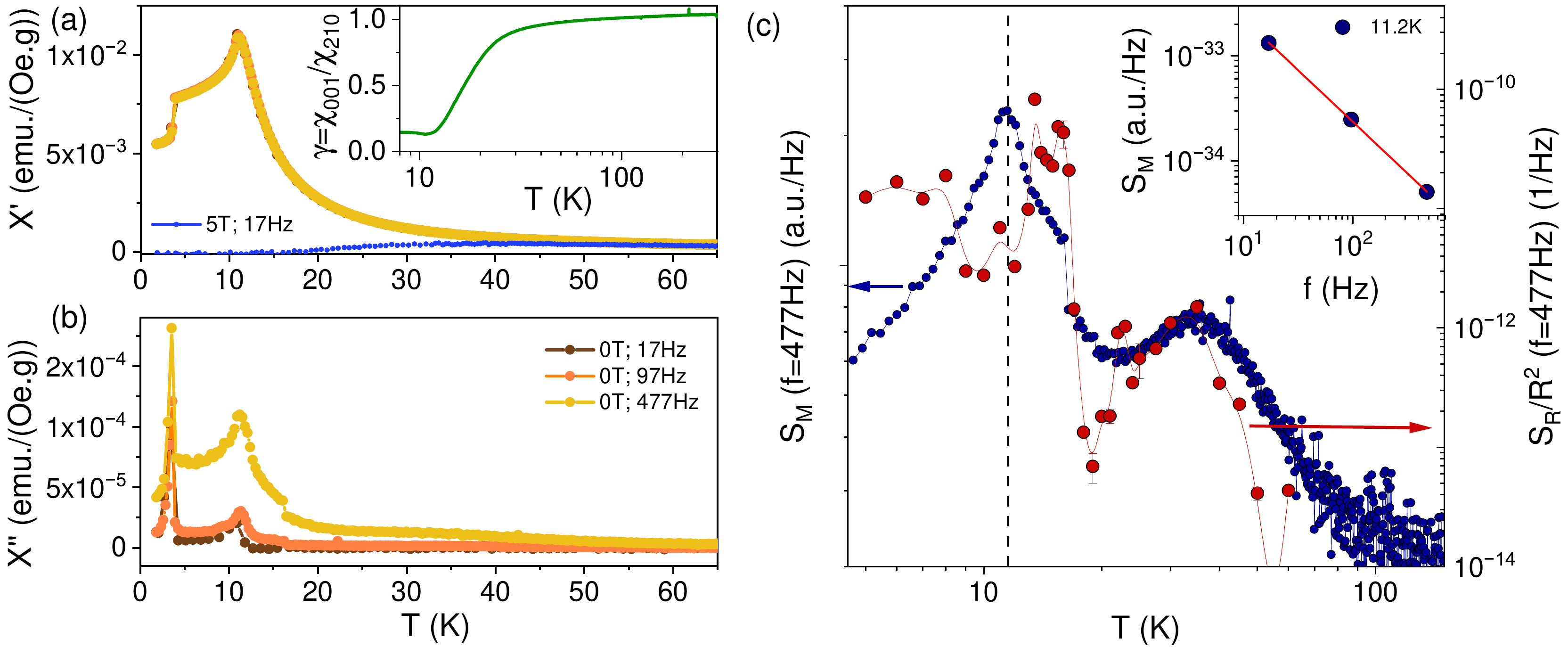}
\caption{\label{fig:magnet_AC}(a) Real part of the AC susceptibility $\chi^\prime$ of sample \#2 at zero dc magnetic field for various frequencies between $f=0.97\,$Hz and $f=477\,$Hz. Data at $17\,$Hz measured in an external dc field of $\mu_0 H=5\,$T are shown in blue. The amplitude of the AC field is $\mu_0H = 0.5\,$mT. Inset shows the anisotropy $\gamma (T) = \chi'_{001}(T)/\chi'_{210}(T)$, i.e. the ratio of the zero dc field susceptibilities with the AC field along the $c$-axis and within the $a$-$a$ plane. (b) Imaginary part $\chi^{\prime\prime}$ with the same colour code for different frequencies.  (c) Comparison of the calculated magnetic noise PSD $S_M(f=477\,$Hz) (blue) and the measured resistance noise PSD $S_R/R^2(f=477\,\textrm{Hz})$ (red). Inset demonstrates the $1/f$ dependence of the magnetic noise at $T=11.2\,$K.} 
\end{figure*}

In order to compare low-frequency resistance and magnetic fluctuations we measured the frequency-dependent AC susceptibility of sample \#2. 
Figure \ref{fig:magnet_AC}(a) shows the real part $\chi^\prime(T,f)$ measured at zero external DC field with an AC field of $5\,$ Oe applied along the $c$-axis for different frequencies (yellow to red colors) in comparison to the data taken at $\mu_0 H=5\,$T dc field (blue) for $f=17\,$Hz. For decreasing temperature the susceptibility shows the expected behavior for an antiferromagnet with an essentially frequency-independent cusp marking the  magnetic transition temperature $T_{\textrm{N}}=11\,$K and further decrease in the AFM phase. The jump at $T=3.7\,$K is due to the superconducting transition of Sn  flux incorporated in to the sample during the crystal growth.
In a magnetic field of 5\,T the susceptibility decreases below about $T=40\,$K reaching zero below $T=17\,$K.\\
The imaginary part of the AC susceptibility $\chi''(T)$ is shown in Fig.\ \ref{fig:magnet_AC}(b) for frequencies $f=17-477\,$Hz. The magnitude of $\chi''(T)$ with a sharp peak at $T_{\rm N} = 11.1\,$K is frequency dependent with larger values for higher frequencies. (The anomaly at $T=3.7\,$K is due to the superconducting transition of Sn flux incorporated in the sample during crystal growth.)\\
In the inset of Fig.\ \ref{fig:magnet_AC}(a) the anisotropy $\gamma(T) = \chi'_{001}(T)/\chi'_{210}(T)$ for the AC field along the $c$-axis (hard axis) and within the $a$-$a$ plane (easy axis) is shown. We find $\gamma(T) \approx 1$ (dotted line) at high temperatures while it starts to gradually decrease upon cooling until a sharp downturn occurs below $T\approx25\,$K, i.e.\ the susceptibility for the field applied in the $a$-$a$ plane is increasing faster and becomes larger than for the $c$-direction. Below the magnetic ordering temperature, the anisotropy in the zero-field magnetic susceptibility saturates at a value of $\gamma \sim 0.14$.\\  
In the inset of Fig.\ \ref{fig:magnet_AC}(c) the calculated $S_M (f)$ at $T=11.2\,$K is shown to follow a $1/f^\alpha$ frequency dependence with $\alpha = 0.98$, allowing a direct comparison to the resistance noise measurements which we present in the discussion chapter.

\end{document}